\documentclass[12pt]{article}
\usepackage{latexsym}
\usepackage{amssymb}
\usepackage{amsmath}
\usepackage{amscd}
\usepackage{amsthm}
\usepackage[left=3.0cm,top=2.75cm,right=2.50cm,bottom=2.00cm]{geometry}
\usepackage[dvips]{graphicx}
\usepackage{hyperref}
\usepackage{textcomp}

\begin{document}
\begin{center}
\Large{\bf{Invariant Bianchi type I models in
$f\left(R,T\right)$ Gravity}}

\vspace{10mm}

\normalsize{Anil Kumar Yadav$^{\dag}$, Ahmad T. Ali$^{\ddag}$} \\
\vspace{2mm}\normalsize{$^\dag$ Department of Physics\\ United College of Engineering and 
Research, Greater Noida - 201310, India.\\
E-mail: abanilyadav@yahoo.co.in}\\
\vspace{2mm}
\normalsize{$^\ddag$ King Abdul Aziz University,
Faculty of Science, Department of Mathematics,\\
PO Box 80203, Jeddah, 21589, Saudi Arabia.}\\
E-mail: atali71@yahoo.com\\
\normalsize{$^\ddag$ Mathematics Department, Faculty of Science, Al-Azhar University,\\
Nasr city, 11884, Cairo, Egypt}
\end{center}
\begin{abstract}
In this paper, we search the existence of invariant solutions of Bianchi
type I space-time in the context of $f\left(R,T\right)$ gravity. The exact solution of 
the Einstein's field equations are derived by using Lie point symmetry 
analysis method that yield two models of invariant universe for symmetries $X^{(1)}$ and $X^{(3)}$. The model 
with symmetries $X^{(1)}$ begins with big bang singularity while the model with symmetries $X^{(3)}$ does not favour the 
big bang singularity. Under this specification, we find out at set of singular and non singular solution of Bianchi 
type I model which present several other physically valid features within the framework of $f\left(R,T\right)$. \\  
\end{abstract}

\emph{PACS:} 98.80.JK, 98.80.-k.

\emph{Keywords}: Invariant solutions, Bianchi type - I,
$f\left(R,T\right)$ Gravity.

\section{Introduction }
To explain the observed late time acceleration of the universe, one may assume 
that at large scale, general relativity breaks down and a more general action 
for the gravitational field needs to be invoked. In this paper, we investigate 
$f(R,T)$ gravity, which is a modification of general relativity in which the geometrical 
part of the Einstein-Hilbert action is different. Apart from the Ricci scalar R in the 
Lagrangian, there is also an arbitrary function of the trace T of the energy momentum tensor. 
This theory has attracted much attention in the recent past, and various aspects of the theory has been 
studied. It is possible to explain the dark energy and the observed late time acceleration of the universe 
within this theory. Harko et al \cite{harko1} have obtained the equation of motion of the test particle 
and gravitational field equations in the metric formalism. Marzakulov \cite{myrzakulov2011} presented the point like 
Lagrangian for $f(R,T)$ gravity. Latter on Houndjo \cite{houndjo2012} considered the $f(R,T)$ gravity model 
that satisfies 
the local tests and transition of matter from the dominated era to the accelerated phase of universe. Recently 
Yadav \cite{yadav2014} has investigated anisotropic string cosmological model in $f(R,T)$ gravity. In this paper, 
our aim is ti study invariant Bianchi type I cosmological models in $f(R,T)$ gravity by using Lie point symmetry 
analysis method. One can refers the detail for Lie point symmetry analysis methods and optimal system in the references 
\cite{ali2014}$-$\cite{ovsi1}. We note that relate to  $f\left(R,T\right)$ gravity in anisotropic space-time, 
there are lot of work 
available in the literature based on different mathematical as well as physical 
issues \cite{ahmad2014}$-$\cite{sahoo2014}. In the present model 
our motivation is to study the invariant model of universe under $f\left(R,T\right)$ gravity and to observe 
different physical features of the model. \\

In the recent years, there has been considerable interest in anisotropic space-time that have consistency with 
CMB observations. Bianchi type I model is the most simplest model of anisotropy universe whose spatial section 
are flat but the expansion rate are direction dependent. In the literature, to study the possible effects of 
anisotropy in the early universe, many authors (Yadav et al \cite{yadav2012a}, Akarsu and Kiinc \cite{akarsu2010}, 
Kumar and Singh \cite{kumar2011}, Yadav and Saha 
\cite{yadav2012}, Yadav \cite{yadav2016}, Sahoo et al \cite{sahoo2017}) have investigated Bianchi type I model from 
different point of view. Recently, Sahoo et al \cite{sahoo2017} have investigated Bianchi type I magnetized 
strange quark model with big rip singularity in $f\left(R,T\right)$ and found that the model begins with big bang and 
ends with big rip. In 2010, Saha and Visinescu \cite{saha2010} and Saha et al \cite{saha2010a} have studied Bianchi type 
I string cosmological model in presence of magnetic flux and the study reveals that the presence of cosmic strings 
do not allow the anisotropic universe into an isotropic one. In this paper, we confine ourselves to study the Bianchi 
type I model in the framework of $f\left(R,T\right)$ with different approach i. e. Lie point symmetry 
analysis method.\\

The outline of present study is therefore as follows: 
In section 2, the basic mathematical formalism of $f\left(R,T\right)$ gravity has been given. 
Thereafter in section 3, we provide the anisotropic metric and field equations for invariant model of 
universe in $f\left(R,T\right)$ gravity. The section 4 include the description of optical system and 
symmetry analysis method that leads the exact solution of the model. The solution of the field equations are given in 
section 5. Several physical properties of the model viz expansion scalar, deceleration parameter and energy density, 
are discussed in section 6. Finally, some concluding remarks are given in section 7.\\

\section{Some Basics of $f\left(R,T\right)$ Gravity}

The $f\left(R,T\right)$ theory of gravity is the generalization or
modification of general relativity (GR). The action for this theory
is given by\cite{harko1, anil2014}: 
\begin{equation}
 \label{FRT1}
S\,=\,\int\,\sqrt{-g}\,\left(\dfrac{f\left(R,T\right)}{16\,\pi\,G}+L_m\right)\,dx^4,
\end{equation}
where $f\left(R,T\right)$ is an arbitrary function of the Ricci
scalar $R$ and the trace $T$ of energy momentum tensor $T_{\mu\nu}$
while $L_m$ is the usual matter Lagrangian. It is worth mentioning
that if we replace $f\left(R,T\right)$ with $f\left(R\right)$ leads
to the action of GR. The energy momentum tensor $T_{\mu\nu}$ is
defined as \cite{landau1}:
\begin{equation}
 \label{FRT2}
T_{\mu\nu}\,=\,-\dfrac{2}{\sqrt{-g}}\,\dfrac{\delta\left(\sqrt{-g}\,L_m\right)}{\delta
g^{\mu\nu}}.
\end{equation}
Here we assume that the dependence of matter Lagrangian is merely on
the metric tensor $g_{\mu\nu}$ rather than its derivatives. In this
case, we obtain
\begin{equation}
 \label{FRT3}
T_{\mu\nu}\,=\,L_m\,g_{\mu\nu}-\delta\,\dfrac{\delta L_m}{\delta
g^{\mu\nu}}.
\end{equation}
The $f\left(R,T\right)$ gravity field equations are obtained by
varying the action $S$ with respect to the metric tensor
$g_{\mu\nu}$
\begin{equation}
 \label{FRT4}
f_R\left(R,T\right)\,R_{\mu\nu}-\dfrac{1}{2}\,f\left(R,T\right)\,g_{\mu\nu}-\left(\nabla_\mu\,\nabla_\nu-g_{\mu\nu}\,\Box\right)f_R\left(R,T\right)
=\kappa\,T_{\mu\nu}-f_T\left(R,T\right)\,\left(T_{\mu\nu}+\Theta_{\mu\nu}\right),
\end{equation}
where $\nabla_\mu$ denotes the covariant derivative and
$\Box=\nabla^\mu\,\nabla_\mu$, $f_R\left(R,T\right)=\dfrac{\partial
f\left(R,T\right)}{\partial R}$,
$f_T\left(R,T\right)=\dfrac{\partial f\left(R,T\right)}{\partial T}$
and $\Theta_{\mu\nu}=g^{\alpha\beta}\,\dfrac{\delta
T_{\alpha\beta}}{\delta g^{\mu\nu}}$. Contraction of (\ref{FRT4})
yields
\begin{equation}
 \label{FRT5}
R\,f_R\left(R,T\right)+3\,\Box\,f_R\left(R,T\right)-2\,f\left(R,T\right)
=\kappa\,T-\left(T+\Theta\right)\,f_T\left(R,T\right),
\end{equation}
where $\Theta=\Theta^\mu_\mu$. This is an important equation because
it provides a relationship between Ricci scalar $R$ and the trace
$T$ of energy momentum tensor. Using matter Lagrangian $L_m$, the
standard matter energy-momentum tensor is derived as
\begin{equation}
 \label{FRT6}
T_{\mu\nu}\,=\,(p+\rho)\,u_\mu\,u_\nu-p\,g_{\mu\nu},
\end{equation}
where $u_\mu=\sqrt{g_{00}}\,\left(1,0,0,0\right)$ is the
four-velocity in co-moving coordinates and $\rho$ and $p$ denotes
energy density and pressure of the fluid respectively. Perfect
fluids problems involving energy density and pressure are not any
easy task to deal with. Moreover, there does not exist any unique
definition for matter Lagrangian. Thus we can assume the matter
Lagrangian as $L_m=-p$ which gives
\begin{equation}
 \label{FRT7}
\Theta_{\mu\nu}\,=\,-p\,g_{\mu\nu}-2\,T_{\mu\nu},
\end{equation}
and consequently the field equation (\ref{FRT4})takes the form
\begin{equation}
 \label{FRT8}
f_R\left(R,T\right)\,R_{\mu\nu}-\dfrac{1}{2}\,f\left(R,T\right)\,g_{\mu\nu}-\left(\nabla_\mu\,\nabla_\nu-g_{\mu\nu}\,\Box\right)f_R\left(R,T\right)
=\kappa\,T_{\mu\nu}+f_T\left(R,T\right)\,\left(T_{\mu\nu}+p\,g_{\mu\nu}\right),
\end{equation}
It is mentioned here that these field equations depend on the
physical nature of matter field. Many theoretical models
corresponding to different matter contributions for
$f\left(R,T\right)$ gravity are possible. However, Harko et. al.
gave three classes of these models
\begin{equation}
 \label{FRT9}
f\left(R,T\right)\,=\,\left\{
                        \begin{array}{ll}
                          R+2\,f\left(T\right),\\
                          f_1\left(R\right)+f_2\left(T\right),\\
                          f_1\left(R\right)+f_2\left(R\right)\,f_3\left(T\right).
                        \end{array}
                      \right.
\end{equation}
In this paper we focused to the first class, i.e.,
$f\left(R,T\right)\,=\,R+2\,f\left(T\right)$. For this model the
field equations become
\begin{equation}
 \label{FRT10}
R_{\mu\nu}-\dfrac{1}{2}\,R\,g_{\mu\nu}\,=\,\Big[\kappa+2\,f'\left(T\right)\Big]\,T_{\mu\nu}+\Big[f\left(T\right)+2\,p\,f'\left(T\right)\Big]\,g_{\mu\nu},
\end{equation}
where prime represents derivative with respect to $T$.

\section{The metric and field equations}
The line element of the
Bianchi type I space-time is given by
\begin{equation}
 \label{spacetime}
ds^{2}=dt^{2}-A^2\,dx^{2}-B^2\,dy^{2}-C^2\,dz^2,
\end{equation}
where $A=A(x,t)$, $B=B(x,t)$ and $C=C(x,t)$ are functions of $x$ and
$t$.

Using Equation (\ref{FRT10}), we get following five independent field
equations

\begin{equation}
 \label{efe1}
\lambda\,p-\left(3\,\lambda+8\,\pi\right)\,\rho=\dfrac{1}{A^{2}}\left[\dfrac{C''}{C}+\dfrac{B'\,C'}{B\,C}+\dfrac{B''}{B}-\dfrac{A'\,C'}{A\,C}-\dfrac{A'\,B'}{A\,B}\right]-
\dfrac{\dot{B}\,\dot{C}}{B\,C}-\dfrac{\dot{A}\,\dot{C}}{A\,C}-\dfrac{\dot{A}\,\dot{B}}{A\,B},
\end{equation}

\begin{equation}
 \label{efe2}
\left(3\,\lambda+8\,\pi\right)\,p-\lambda\,\rho=\dfrac{B'\,C'}{A^2\,B\,C}-
\dfrac{\ddot{C}}{C}-\dfrac{\dot{B}\,\dot{C}}{B\,C}-\dfrac{\ddot{B}}{B},
\end{equation}

\begin{equation}
 \label{efe3}
\left(3\,\lambda+8\,\pi\right)\,p-\lambda\,\rho=\dfrac{1}{A^2}\left[\dfrac{C''}{C}-\dfrac{A'\,C'}{A\,C}\right]-
\dfrac{\ddot{C}}{C}-\dfrac{\dot{A}\,\dot{C}}{A\,C}-\dfrac{\ddot{A}}{A},
\end{equation}

\begin{equation}
 \label{efe4}
\left(3\,\lambda+8\,\pi\right)\,p-\lambda\,\rho=\dfrac{1}{A^2}\left[\dfrac{B''}{B}-\dfrac{A'\,B'}{A\,B}\right]-
\dfrac{\ddot{B}}{B}-\dfrac{\dot{A}\,\dot{B}}{A\,B}-\dfrac{\ddot{A}}{A},
\end{equation}

\begin{equation}
 \label{efe5}
\dfrac{\dot{B}^{\prime}}{B}+\dfrac{\dot{C}^{\prime}}{C}=\dfrac{\dot{A}}{A}\left(\dfrac{B^{\prime}}{B}+\dfrac{C^{\prime}}{C}\right).
\end{equation}
Here $A^{\prime}=\dfrac{dA}{dx}$, $\dot{A} = \dfrac{dA}{dt}$, $G = c = 1$ and $f(T) = \lambda T$.

The scalar expansion
$(\Theta)$, shear scalar $(\sigma^2)$
and proper volume $V$ are respectively obtained from following well known expressions\cite{dec1}:

\begin{equation}  \label{u215}
\Theta\,=\,u_{;i}^{i}=\dfrac{\dot{A}}{A}+\dfrac{\dot{B}}{B}+\dfrac{\dot{C}}{C},
\end{equation}

\begin{equation}  \label{u216}
\begin{array}{ll}
\sigma^2\,=\,\dfrac{1}{2}\,\sigma_{ij}\,\sigma^{ij}=\dfrac{\Theta^2}{3}-
\dfrac{\dot{A}\dot{B}}{A\,B}-\dfrac{\dot{A}\,\dot{C}}{A\,C}-\dfrac{\dot{B}\,\dot{C}}{B\,C},
\end{array}
\end{equation}

\begin{equation}  \label{u217}
\dot{u}_i\,=\,u_{i;j}\,u^j\,=\,\big(0,0,0,0\big),
\end{equation}

\begin{equation}  \label{u218}
V=\sqrt{-g}=A\,B\,C,
\end{equation}
where $g$ is the determinant of the metric (\ref{spacetime}).\\
The shear tensor is
\begin{equation}  \label{u219}
  \begin{array}{ll}
\sigma_{ij}\,=\,u_{(i;j)}+\dot{u}_{(i}\,u_{j)}-\frac{1}{3}\,\Theta\,(g_{ij}-u_i\,u_j)
\,=\,u_{i;j}+\frac{1}{2}\left(u_{i;k}\,u^k\,u_j+u_{j;k}\,u^k\,u_i\right)-\frac{1}{3}\,\Theta\,(g_{ij}-u_i\,u_j).
\end{array}
\end{equation}
and the non-vanishing components of the $\sigma_i^j$ are
\begin{equation}  \label{u220}
\left\{
  \begin{array}{ll}
    \sigma_0^0\,=\,0,\,\,\,\,\,\,\,\,\,\,\,\,\,\,\,\,\,\,\,\,\,\,\,\,\,\,\,\,\,
\,\,\,\,\,\,\,\,\,\,\,\,\,\,\,\,\,\,\,\,\,\,\,\,\,\,\,\,\,\,
\sigma_1^1\,=\,\dfrac{1}{3}\left(\dfrac{2\,\dot{A}}{A}-\dfrac{\dot{B}}{B}-\dfrac{\dot{C}}{C}\right),\\
\\
\sigma_2^2\,=\,\dfrac{1}{3}\left(\dfrac{2\,\dot{B}}{B}-\dfrac{\dot{C}}{C}-\dfrac{\dot{A}}{A}\right),\,\,\,\,\,\,\,\,\,\,
\sigma_3^3\,=\dfrac{1}{3}\left(\dfrac{2\,\dot{C}}{C}-\dfrac{\dot{A}}{A}-\dfrac{\dot{B}}{B}\right).
   \end{array}
\right.
\end{equation}

The field equations (\ref{efe1})-(\ref{efe5}) constitute a system of
five highly non-linear partial differential equations (NLPDE) with five unknowns parameters namely, $A$, $B$, $C$,
$\rho$ and $p$.  The equations (\ref{efe1})-(\ref{efe5}) can be
transform to the following NLPDE.
\begin{equation}  \label{u210-1}
\begin{array}{ll}
    E_1=\dfrac{1}{A^2}\left[\dfrac{B''}{B}-\dfrac{B'\,C'}{B\,C}-\dfrac{A'\,B'}{A\,B}\right]+\dfrac{\ddot{C}}{C}-\dfrac{\ddot{A}}{A}+\dfrac{\dot{B}\,\dot{C}}{B\,C}
    -\dfrac{\dot{A}\,\dot{B}}{A\,B}=0,
  \end{array}
\end{equation}

\begin{equation}  \label{u210-2}
\begin{array}{ll}
        E_2=\dfrac{1}{A^2}\left[\dfrac{C''}{C}-\dfrac{B''}{B}-\dfrac{A'\,C'}{A\,C}+\dfrac{A'\,B'}{A\,B}\right]
        +\dfrac{\ddot{B}}{B}-\dfrac{\ddot{C}}{C}-\dfrac{\dot{A}\,\dot{C}}{A\,C}+\dfrac{\dot{A}\,\dot{B}}{A\,B}=0,
  \end{array}
\end{equation}

\begin{equation}  \label{u210-3}
\begin{array}{ll}
        E_3=\dfrac{\dot{B}^{\prime}}{B}+\dfrac{\dot{C}^{\prime}}{C}-\dfrac{\dot{A}}{A}\left(\dfrac{B^{\prime}}{B}+\dfrac{C^{\prime}}{C}\right)=0,
  \end{array}
\end{equation}
where
\begin{equation}  \label{u210-4}
\begin{array}{ll}
        8\,(4\,\pi+\lambda)\,(2\,\pi+\lambda)\,p(x,t)\,=\lambda\,\Bigg[\dfrac{\dot{A}\,\dot{C}}{A\,C}+\dfrac{\dot{B}\,\dot{C}}{B\,C}
-\dfrac{1}{A^2}\,\left(\dfrac{C''}{C}+\dfrac{B'\,C'}{B\,C}-\dfrac{A'\,C'}{A\,C}\right)
        \Bigg]\\
        \\
        \,\,\,\,\,\,\,\,\,\,\,\,\,\,\,\,\,\,\,\,\,\,\,\,\,\,\,\,\,\,\,\,\,\,\,\,\,\,\,\,\,\,\,\,\,\,\,\,\,\,\,\,\,\,\,\,\,\,\,\,
-(8\,\pi+2\,\lambda)\,\Bigg[\dfrac{\dot{A}\,\dot{B}}{A\,B}+\dfrac{1}{A^2}\,\left(\dfrac{A'\,B'}{A\,B}-\dfrac{B''}{B}\right)\Bigg]
-(8\,\pi+3\,\lambda)\,\left(\dfrac{\ddot{A}}{A}+\dfrac{\ddot{B}}{B}\right),
  \end{array}
\end{equation}

\begin{equation}  \label{u210-5}
\begin{array}{ll}
        8\,(4\,\pi+\lambda)\,(2\,\pi+\lambda)\,\rho(x,t)\,=(8\,\pi+3\,\lambda)\,\Bigg[\dfrac{\dot{A}\,\dot{C}}{A\,C}+\dfrac{\dot{B}\,\dot{C}}{B\,C}
-\dfrac{1}{A^2}\,\left(\dfrac{C''}{C}+\dfrac{B'\,C'}{B\,C}-\dfrac{A'\,C'}{A\,C}\right)
        \Bigg]\\
        \\
        \,\,\,\,\,\,\,\,\,\,\,\,\,\,\,\,\,\,\,\,\,\,\,\,\,\,\,\,\,\,\,\,\,\,\,\,\,\,\,\,\,\,\,\,\,\,\,\,\,\,\,\,\,\,\,\,\,\,\,\,
+(8\,\pi+2\,\lambda)\,\Bigg[\dfrac{\dot{A}\,\dot{B}}{A\,B}+\dfrac{1}{A^2}\,\left(\dfrac{A'\,B'}{A\,B}-\dfrac{B''}{B}\right)\Bigg]
-\lambda\,\left(\dfrac{\ddot{A}}{A}+\dfrac{\ddot{B}}{B}\right).
  \end{array}
\end{equation}
\section{Optimal system and Symmetry analysis method}
The classical method
for finding the solution of equations (\ref{u210-1})-(\ref{u210-5}) is a separation method by taking
$A(x,t)=A_1(x)\,A_2(t)$, $B(x,t)=B_1(x)\,B_2(t)$ and
$C(x,t)=C_1(x)\,C_2(t)$ \cite{pradhan2007,yadav2009}. In order to obtain an exact and 
new solution of equations (\ref{u210-1})-(\ref{u210-5}), here we have used the 
symmetry analysis method which is a powerful method for solving NLPDE. In refs. 
\cite{anil2014}$-$\cite{yadav2015}, the authors have described optimal system and the symmetry analysis method 
Bianchi-I in detail. Following, Yadav and Ali \cite{anil2014} and Ali et al \cite{yadav2015}, we acquire 
an optimal system of one-dimensional subalgebras which is spanned
as following
\begin{equation}\label{u32-5}
\begin{array}{ll}
\{X^{(1)}=X_1+a_3\,X_3+a_5\,X_5+a_6\,X_6,\,\,\,\,\,\,X^{(2)}=X_1+a_4\,X_4+a_5\,X_5+a_6\,X_6,\,\\
\,\,\,X^{(3)}=X_2+a_3\,X_3+a_5\,X_5+a_6\,X_6,\,\,\,\,\,\,X^{(4)}=X_2+a_4\,X_4+a_5\,X_5+a_6\,X_6,\,\\
\,\,\,X^{(5)}=X_3+a_5\,X_5+a_6\,X_6,\,\,\,\,\,X^{(6)}=X_4+a_5\,X_5+a_6\,X_6,\,\,\,\,\,
X^{(7)}=X_5+a_6\,X_6,\,\,\,\,\,X^{(8)}=X_6\}.
\end{array}
\end{equation}   

\section{Invariant solutions}

In the case of symmetries $X^{(5)}$ or $X^{(6)}$ or $X^{(7)}$ or $X^{(8)}$, then $a_1=a_2=0$ that leads 
the metric functions $A$, $B$ and $C$ are functions of $x$ only.
For this reason, we shall consider the invariant solutions associated with the optimal systems of 
symmetries $X^{(1)}$, $X^{(2)}$, $X^{(3)}$ and $X^{(4)}$ only. 
Among these symmetries, only $X^{(1)}$ gives the model of physical universe while the model with 
$X^{(3)}$ does not favour the big bang singularity. In this paper, we have purposely 
omitted the models with symmetries $X^{(2)}$ and $X^{(4)}$ because this prescription represent non-realistic 
model of universe.\\

\textbf{Solution (I):} The symmetries $X^{(1)}$ has the
characteristic equations:
\begin{equation}\label{u41-1}
\dfrac{dx}{x}=\dfrac{dt}{a_3\,t}=\dfrac{dA}{(a_3-1)\,A}=\dfrac{dB}{a_5\,B}=\dfrac{dC}{a_6\,C}.
\end{equation}
Then the Invariant transformations take the following form:
\begin{equation}\label{u42-1}
\begin{array}{ll}
\xi=\dfrac{x^a}{t},\,\,\,\,\,\,A(x,t)=\,x^{a-1}\,\Psi(\xi),\,\,\,\,\,\,B(x,t)=\,x^{b}\,\Phi(\xi),\,\,\,\,\,\,C(x,t)=\,x^{c}\,\Omega(\xi),
\end{array}
\end{equation}
where $a=a_3$, $b=a_5$ and $c=a_6$ are an arbitrary constants.\\
Putting the transformations (\ref{u42-1}) in the field Eqs.
(\ref{u210-1})-(\ref{u210-3}), we have
\begin{equation}\label{u43-1}
\begin{array}{ll}
a\,\xi\,\left(\dfrac{\Omega''}{\Omega}+\dfrac{\Phi''}{\Phi}-\dfrac{\Psi'\,\Omega'}{\Psi\,\Omega}-\dfrac{\Psi'\,\Phi'}{\Psi\,\Phi}\right]
=(b+c)\,\dfrac{\Psi'}{\Psi}-(a+b)\,\dfrac{\Phi'}{\Phi}-(a+c)\,\dfrac{\Omega'}{\Omega},
\end{array}
\end{equation}

\begin{equation}\label{u43-2}
\begin{array}{ll}
\xi^4\,\left(\dfrac{\Psi''}{\Psi}-\dfrac{\Omega''}{\Omega}+\dfrac{\Psi'\,\Phi'}{\Psi\,\Phi}-\dfrac{\Phi'\,\Omega'}{\Phi\,\Omega}\right)
+2\,\xi^3\,\left(\dfrac{\Psi'}{\Psi}-\dfrac{\Omega'}{\Omega}\right)
+\dfrac{a^2\,\xi^2}{\Psi^2}\left(\dfrac{\Psi'\,\Phi'}{\Psi\,\Phi}+\dfrac{\Phi'\,\Omega'}{\Phi\,\Omega}-\dfrac{\Phi''}{\Phi}\right)\\
\\
\,\,\,\,\,\,\,\,\,\,\,\,\,\,\,\,\,\,\,\,\,\,\,\,\,\,\,\,\,\,\,\,\,\,\,\,\,\,\,\,\,\,\,\,\,\,\,\,\,\,
+\dfrac{a\,\xi}{\Psi^2}\,\Bigg[b\,\left(\dfrac{\Psi'}{\Psi}+\dfrac{\Omega'}{\Omega}\right)+(c-2\,b)\,\dfrac{\Phi'}{\Phi}\Bigg]=\dfrac{b\,(b-a-c)}{\Psi^2},
\end{array}
\end{equation}

\begin{equation}\label{u43-3}
\begin{array}{ll}
\xi^4\,\left(\dfrac{\Phi''}{\Phi}-\dfrac{\Omega''}{\Omega}+\dfrac{\Psi'\,\Phi'}{\Psi\,\Phi}-\dfrac{\Psi'\,\Omega'}{\Psi\,\Omega}\right)
+2\,\xi^3\,\left(\dfrac{\Phi'}{\Phi}-\dfrac{\Omega'}{\Omega}\right)
+\dfrac{a^2\,\xi^2}{\Psi^2}\left(\dfrac{\Omega''}{\Omega}-\dfrac{\Phi''}{\Phi}+\dfrac{\Psi'\,\Phi'}{\Psi\,\Phi}-\dfrac{\Psi'\,\Omega'}{\Psi\,\Omega}\right)\\
\\
\,\,\,\,\,\,\,\,\,\,\,\,\,\,\,\,\,\,\,\,\,\,\,\,\,\,\,\,\,\,\,\,\,\,\,\,\,\,\,\,\,\,\,\,\,\,\,\,\,\,
+\dfrac{a\,\xi}{\Psi^2}\,\Bigg[(b-a)\,\dfrac{\Psi'}{\Psi}+2\,c\,\dfrac{\Omega'}{\Omega}+2\,b\,\dfrac{\Phi'}{\Phi}\Bigg]=\dfrac{(c-b)\,(b-a-c)}{\Psi^2},
\end{array}
\end{equation}

One can not solve equations (\ref{u43-1})-(\ref{u43-3}) in general. So, in 
order to solve the problem completely, we have to choose the following transformations:
\begin{equation}\label{u43-4}
\begin{array}{ll}
\Psi(\xi)=\alpha_1\,\xi^{\alpha_2},\,\,\,\,\,\,\,\,\,\,\Phi(\xi)=\beta_1\,\xi^{\beta_2},\,\,\,\,\,\,\,\,\,\,\Omega(\xi)=\gamma_0+\gamma_1\,\xi^{\gamma_2},
\end{array}
\end{equation}
where $\alpha_1$, $\alpha_2$, $\beta_1$, $\beta_2$, $\gamma_0$,
$\gamma_1$ and $\gamma_2$ are arbitrary non-zero constants.\\
Substituting eq. (\ref{u43-4}) in eq. (\ref{u43-1}), we have

\begin{equation}\label{u43-7}
\begin{array}{ll}
\gamma_0\,\Big[c\,\alpha_2+(b+a\,\beta_2)\,(\alpha_2-\beta_2)\Big]
+\gamma_1\,\Big[(b+a\,\beta_2)\,(\alpha_2-\beta_2)+(c+a\,\gamma_2)\,(\alpha_2-\gamma_2)\Big]\,\xi^{\gamma_2}=0.
\end{array}
\end{equation}
The coefficients of $\xi^{\gamma_2}$ and the absolute value must be
equal zero. Solving the two resulting conditions with respect to $a$
and $b$, we have:
\begin{equation}\label{u43-8}
\begin{array}{ll}
a=\dfrac{c}{\alpha_2-\gamma_2},\,\,\,\,\,\,\,\,\,\,\,\,\,
b=\dfrac{\alpha_2\,c}{\beta_2-\alpha_2}-\dfrac{\beta_2\,c}{\gamma_2-\alpha_2}.
\end{array}
\end{equation}
where $\gamma_2\neq\alpha_2$ and $\beta_2\neq\alpha_2$.\\
Therefore
the (\ref{u43-2}) can be written as
\begin{equation}\label{u43-9}
\begin{array}{ll}
Q_0+Q_1\,\xi^{2+2\,\alpha_2}+Q_2\,\xi^{\gamma_4}+Q_3\,\xi^{2+2\,\alpha_2+\gamma_2}\,=\,0,
\end{array}
\end{equation}
where
\begin{equation}\label{u43-10}
\left\{
  \begin{array}{ll}
   Q_0=\dfrac{c^2\,\alpha_2\,\gamma_0}{(\alpha_2-\beta_2)^2\,(\alpha_2-\gamma_2)}\,\Big[
\beta_2\,(1-\gamma_2)+\alpha_2\,(2\,\beta_2+2\,\gamma_2-3\,\alpha_2-1)\Big],\\
\\
Q_1=\alpha_1^2\,\alpha_2\,\gamma_0\,(1+\alpha_2+\beta_2),\\
\\
Q_2=\dfrac{c^2\,\alpha_2\,\gamma_1}{(\alpha_2-\beta_2)^2\,(\alpha_2-\gamma_2)}\,\Big[
\beta_2+\alpha_2\,(2\,\beta_2+\gamma_2-3\,\alpha_2-1)\Big],\\
\\
Q_3=\alpha_1^2\,\gamma_1\,(\alpha_2-\gamma_2)\,(1+\alpha_2+\beta_2+\gamma_2).
  \end{array}
\right.
\end{equation}
From equation (\ref{u43-9}), we have two cases: \textbf{(1):}
$2+2\,\alpha_2=0$ and \textbf{(2):} $2+2\,\alpha_2\neq0$.\\
In case (2), all $Q_i,\,i=0,1,2,3$ must be equal to zero. The equation $Q_1=0$
leads to $\beta_2=-\alpha_2-1$ and the coefficient $Q_4$ becomes
$\alpha_1^2\,\gamma_1\,(\alpha_2-\gamma_2)\,\gamma_2$ which equal
zero when $\gamma_2=\alpha_2$ contradiction.\\
However, in case (1), i.e. $\alpha_2=-1$.\\
The equation (\ref{u43-9}) transform to the following form:
\begin{equation}\label{u43-10-0}
\begin{array}{ll}
\gamma_0\,(\gamma_2+1)\,\Big[\alpha_1^2\,\beta_2\,(\beta_2+1)^2+c^2\,(\beta_2+2)\Big]
+\gamma_1\,\Big[\alpha_1^2\,(\beta_2+1)^2\,(\gamma_2+1)^2+c^2\,(\gamma_2+\beta_2+2)\Big]\,\xi^{\gamma_2}=0.
\end{array}
\end{equation}
The equation (\ref{u43-10-0}) leads to
\begin{equation}\label{u43-8}
\begin{array}{ll}
c=\pm\,\dfrac{\alpha_1\,\delta_0\,(1-\delta_0^2)}{1+\delta_0^2},\,\,\,\,\,\,\,\,\,\,\,\,\,
\beta_2=\dfrac{-1\pm\delta_0\,\sqrt{1+\delta_0^2-\delta_0^4}}{1+\delta_0^2},
\end{array}
\end{equation}
where $\gamma_2=-\dfrac{2\,\delta_0^2}{1+\delta_0^2}$.\\
Thus, the metric functions are given by
\begin{equation}\label{uu1}
A(x,t)=\dfrac{\alpha_1\,t}{x},\,\,\,\,\,
B(x,t)=\beta_1\,x^{\frac{\alpha_1\,(1-\delta_0^2)}{\delta_0+\sqrt{1+\delta_0^2-\delta_0^4}}}\,t^{\frac{1-\delta_0\,\sqrt{1+\delta_0^2-\delta_0^4}}{1+\delta_0^2}},\,\,\,\,\,
C(x,t)=\Bigg(\gamma_1\,x^{\frac{2\,\alpha_1\,\delta_0^3}{1+\delta_0^2}}+\gamma_2\,t^{\frac{2\,\delta_0^2}{1+\delta_0^2}}\Bigg)\,x^{-\alpha_1\,\delta_0},
\end{equation}
 The equations (\ref{uu1}) and (\ref{spacetime}) lead to
\begin{equation}  \label{s1}
\begin{array}{ll}
ds_{1}^2=dt^2-\dfrac{\alpha_1^2\,t^2}{x^2}\,dx^2+
\beta_1^2\,x^{\frac{2\,\alpha_1\,(1-\delta_0^2)}{\delta_0+\sqrt{1+\delta_0^2-\delta_0^4}}}\,t^{\frac{2-2\delta_0\,\sqrt{1+\delta_0^2-\delta_0^4}}{1+\delta_0^2}}\,dy^2
+\Bigg(\gamma_0\,x^{\frac{2\,\alpha_1\,\delta_0^3}{1+\delta_0^2}}+\gamma_1\,t^{\frac{2\,\delta_0^2}{1+\delta_0^2}}\Bigg)^2\,x^{-2\alpha_1\,\delta_0}\,dz^2.
\end{array}
\end{equation}
\textbf{Solution (II):} The symmetries $X^{(3)}$ has the
characteristic equations:
\begin{equation}\label{u61-1}
\dfrac{dx}{1}=\dfrac{dt}{a_3\,t}=\dfrac{dA}{a_3\,A}=\dfrac{dB}{a_5\,B}=\dfrac{dC}{a_6\,C}.
\end{equation}
Then the Invariant transformations take the following form:
\begin{equation}\label{u62-1}
\begin{array}{ll}
\xi=t\,\exp\big[a\,x\big],\,\,\,\,\,\,A(x,t)=\Psi(\xi)\,t,\,\,\,\,\,\,B(x,t)=\Phi(\xi)\,t^b,\,\,\,\,\,\,C(x,t)=\Omega(\xi)\,t^c,
\end{array}
\end{equation}
where $a=-\frac{1}{a_3}$, $b=\dfrac{a_5}{a_3}$ and
$c=\dfrac{a_6}{a_3}$ are arbitrary constants.\\ 
Putting the transformations (\ref{u62-1}) in the field Equations
(\ref{u210-1})-(\ref{u210-3}), we can get the following system of
ordinary differential equations:
\begin{equation}\label{u63-1}
\begin{array}{ll}
\xi\left(\dfrac{\Omega''}{\Omega}+\dfrac{\Phi''}{\Phi}-\dfrac{\Psi'\,\Phi'}{\Psi\,\Phi}-\dfrac{\Psi'\,\Omega'}{\Psi\,\Omega}\right)
+b\,\dfrac{\Phi'}{\Phi}+c\,\dfrac{\Omega'}{\Omega}=0,
\end{array}
\end{equation}

\begin{equation}\label{u63-2}
\begin{array}{ll}
a^2\,\xi\,\Bigg[\xi\,\left(\dfrac{\Omega''}{\Omega}-\dfrac{\Psi''}{\Psi}+\dfrac{\Phi'\,\Omega'}{\Phi\,\Omega}
-\dfrac{\Psi'\,\Phi'}{\Psi\,\Phi}\right)+
(b+2\,c)\,\dfrac{\Omega'}{\Omega}+(c-1)\,\dfrac{\Phi'}{\Phi}+(b+2)\,\dfrac{\Psi'}{\Psi}\Bigg]
\\
\\
\,\,\,\,\,\,\,\,\,\,\,\,\,\,\,\,\,\,\,\,\,\,\,\,\,\,\,\,\,\,\,\,\,\,\,\,\,\,\,\,\,\,\,\,\,\,\,\,\,\,
\,\,\,\,\,\,\,\,\,\,\,\,\,\,\, +\dfrac{\xi}{\Psi^2}\Bigg[\xi\,\Big(
\dfrac{\Phi''}{\Phi}-\dfrac{\Phi'\,\Omega'}{\Phi\,\Omega}+\dfrac{\Psi'\,\Phi'}{\Psi\,\Phi}\Big)-\dfrac{\Phi'}{\Phi}\Bigg]=\dfrac{(1-c)\,(b+c)}{\Psi^2},
\end{array}
\end{equation}

\begin{equation}\label{u63-3}
\begin{array}{ll}
\xi\,\Bigg[\xi\,\left(\dfrac{\Phi''}{\Phi}-\dfrac{\Omega''}{\Omega}+\dfrac{\Psi'\,\Phi'}{\Psi\,\Phi}
-\dfrac{\Psi'\,\Omega'}{\Psi\,\Omega}\right)-
(1+2\,c)\,\dfrac{\Omega'}{\Omega}+(1+2\,b)\,\dfrac{\Phi'}{\Phi}+(b+c)\,\dfrac{\Psi'}{\Psi}\Bigg]
\\
\\
\,\,\,\,\,\,\,\,\,\,\,\,\,\,\,\,\,\,\,\,\,\,\,\,\,\,\,\,\,\,\,\,\,\,\,\,\,\,\,\,\,\,\,\,\,\,\,\,\,\,
\,\,\,\,\,\,\,\,\,\,\,\,\,\,\,
+\dfrac{a^2\,\xi}{\Psi^2}\Bigg[\xi\,\Big(
\dfrac{\Omega''}{\Omega}-\dfrac{\Omega''}{\Omega}+\dfrac{\Psi'\,\Phi'}{\Psi\,\Phi}-\dfrac{\Psi'\,\Omega'}{\Psi\,\Omega}\Big)
+\dfrac{\Omega'}{\Omega}-\dfrac{\Phi'}{\Phi}\Bigg]=\dfrac{c^2-b^2}{\Psi^2}.
\end{array}
\end{equation}

One can not solve equations (\ref{u43-1})-(\ref{u43-3}) in general. 
So, in order to solve the problem completely, we have to choose the following transformations:
\begin{equation}\label{u63-4}
\begin{array}{ll}
\Psi(\xi)=\alpha_1\,\xi^{\alpha_2},\,\,\,\,\,\,\,\,\,\,\Phi(\xi)=\beta_1\,\xi^{\beta_2},\,\,\,\,\,\,\,\,\,\,\Omega(\xi)=\gamma_0+\gamma_1\,\xi^{\gamma_2},
\end{array}
\end{equation}
where $\alpha_1$, $\alpha_2$, $\beta_1$, $\beta_2$, $\gamma_0$,
$\gamma_1$ and $\gamma_2$ are arbitrary non-zero constants.
Substituting (\ref{u63-4}) in (\ref{u63-1}), we have
\begin{equation}\label{u63-7}
\begin{array}{ll}
\gamma_0\,\left(\beta_2-\alpha_2+b-1\right)
+\gamma_1\,\Big[\beta_2\,\left(\beta_2-\alpha_2+b-1\right)+\gamma_2\,\left(\gamma_2-\alpha_2+c-1\right)\Big]\,\xi^{\gamma_2}=0.
\end{array}
\end{equation}
The coefficients of $\xi^{\gamma_2}$ and the absolute value must be
equal zero in the above equation. Solving the two resulting
conditions with respect to $b$ and $c$, we obtain
\begin{equation}\label{u63-8}
\begin{array}{ll}
b=1+\alpha_2-\beta_2,\,\,\,\,\,\,\,\,\,\,\,\,\,
c=1+\alpha_2-\gamma_2.
\end{array}
\end{equation}
Therefore the (\ref{u63-2}) becomes
\begin{equation}\label{u63-9}
\begin{array}{ll}
\gamma_0\,\beta_2\,a^2\,(\alpha_2-\beta_2)+\gamma_0\,\gamma_2\,\alpha_1^2\,(2+3\,\alpha_2-\gamma_2)\,\xi^{2\,\alpha_2}
+\gamma_1\,\beta_2\,a^2\,(\alpha_2-\beta_2+\gamma_2)\,\xi^{\gamma_2}\,=\,0.
\end{array}
\end{equation}
Because $\alpha_2\neq0$ and $\gamma_2\neq0$, then the absolute value
in the above equation must be equal zero we get
\begin{equation}\label{u63-9-0}
\begin{array}{ll}
\alpha_2\,=\,\beta_2,
\end{array}
\end{equation}
and the equation (\ref{u63-9}) becomes:
\begin{equation}\label{u63-10}
\begin{array}{ll}
\gamma_0\,\alpha_1^2\,(2+3\,\beta_2-\gamma_2)\,\xi^{2\,\beta_2}
+\gamma_1\,\beta_2\,a^2\,\xi^{\gamma_2}\,=\,0.
\end{array}
\end{equation}
From equation (\ref{u63-10}), we have two cases: \textbf{(1):}
$\gamma_2=2\,\beta_2$ and \textbf{(2):} $\gamma_2\neq2\,\beta_2$.
The case (2) leads contradiction while case (1) leads to
$\gamma_0=-\dfrac{\gamma_1\,\beta_2\,a^2}{(2+\beta_2)\,\alpha_1^2}$.
The equations (\ref{u63-4}), (\ref{u62-1}) and (\ref{u43-1})-(\ref{u43-3}) lead to

\begin{equation}\label{uu3}
\left\{
  \begin{array}{ll}
   A(x,t)=\alpha_1\,t^{1+\beta_2}\,\exp[\tilde{a}\,x],\\
\\
B(x,t)=\beta_1\,t^{1+\beta_2}\,\exp[\tilde{a}\,x],\\
\\
C(x,t)=\tilde{\gamma}_1\,\Big[\beta_2\,(2+\beta_2)\,\alpha_1^2\,t^{1+\beta_2}\,\exp[2\,\tilde{a}\,x]-\tilde{a}^2\,t^{1-\beta_2}\Big],
  \end{array}
\right.
\end{equation}
where $\tilde{a}=a\,\beta_2$,
$\tilde{\gamma}_1=\dfrac{\gamma_1}{\beta_2\,(2+\beta_2)\,\alpha_1^2}$,
$\alpha_1$, $\beta_1$ and $\beta_2$ are  arbitrary constants.\\ 
Thus, the line element
(\ref{spacetime}) can be written in the following form:
\begin{equation}  \label{s3}
\begin{array}{ll}
ds_{3}^2=dt^2-t^{2\,(1+\beta_2)}\,\exp[2\,\tilde{a}\,x]\,\Big[\alpha_1^2\,dx^2+\beta_1^2\,dy^2\Big]
+\tilde{\gamma}_1^2\,\Big[\beta_2\,(2+\beta_2)\,\alpha_1^2\,t^{1+\beta_2}\,\exp[2\,\tilde{a}\,x]-\tilde{a}^2\,t^{1-\beta_2}\Big]^2\,dz^2.
\end{array}
\end{equation}
\section{Physical and geometrical properties of the models}

\textbf{For the Model (\ref{s1}):}\\

For the model (\ref{s1}), when $\delta_0=\dfrac{1}{2}$, the
expressions of $p$ and $\rho$ are given by:
\begin{equation}\label{uu1-1}
  \begin{array}{ll}
p(x,t)=\dfrac{9\,\gamma_1\,\Big[(23-4\,\sqrt{19})\,\kappa+(86-28\,\sqrt{19})\lambda\Big]\,x^{\frac{\alpha_1}{5}}
-3\,\gamma_2\,\Big[3\,(4\,\sqrt{19}-23)\,\kappa+2\,(74\,\sqrt{19}-313)\lambda\Big]\,t^{\frac{2}{5}}}{
100\,(\sqrt{19}-2)^2\,(\kappa^2+6\,\kappa\,\lambda+8\,\lambda^2)\,t^2\,\Big(\gamma_1\,x^{\frac{\alpha_1}{5}}+\gamma_2\,t^{\frac{2}{5}}\Big)},
  \end{array}
\end{equation}

\begin{equation}\label{uu1-2}
  \begin{array}{ll}
\rho(x,t)=\dfrac{9\gamma_1\Big[(17-16\sqrt{19})\kappa+(74-55\sqrt{19})\lambda\Big]\,x^{\frac{\alpha_1}{5}}
-3\gamma_2\Big[(112\sqrt{19}-419)\kappa+6(58\sqrt{19}-221)\lambda\Big]\,t^{\frac{2}{5}}}{
100\,(\sqrt{19}-2)^2\,(\kappa^2+6\,\kappa\,\lambda+8\,\lambda^2)\,t^2\,\Big(\gamma_1\,x^{\frac{\alpha_1}{5}}+\gamma_2\,t^{\frac{2}{5}}\Big)},
  \end{array}
\end{equation}
where $\gamma_1$, $\gamma_2$, $\alpha_1$ and $\lambda$ are an
arbitrary constants while $\kappa=8\,\pi$.\\
The volume of universe is read as
\begin{equation}  \label{uu1-4}
V=\alpha_1\,\beta_1\,x^{-1-\frac{(9+2\,\sqrt{19})\,\alpha_1}{10}}\,t^{\frac{18+\sqrt{19}}{10}}\,
\Big(\gamma_1\,x^{\frac{\alpha_1}{5}}+\gamma_2\,t^{\frac{2}{5}}\Big),
\end{equation}
where $\beta_1$ is an arbitrary constant.\\
The expansion scalar,
which determines the volume behavior of the fluid, is given by:
\begin{equation}\label{uu1-5}
  \begin{array}{ll}
\Theta=\dfrac{18+\sqrt{19}}{10\,t}+\dfrac{2\,\gamma_2}{5\,t^{\frac{3}{5}}\,\left(\gamma_1\,x^{\frac{\alpha_1}{5}}+\gamma_2\,t^{\frac{2}{5}}\right)},
  \end{array}
\end{equation}
The non-vanishing components of the shear tensor, $\sigma_i^j$, are read as
\begin{equation}\label{uu1-6}
\left\{
  \begin{array}{ll}
\sigma_0^0\,=\,0,\\
\\
\sigma_1^1\,=\,\dfrac{\gamma_1\,(12-\sqrt{19})\,x^{\frac{\alpha_1}{5}}+\gamma_2\,(8-\sqrt{19})\,t^{\frac{2}{5}}}{
30\,t\,\left(\gamma_1\,x^{\frac{\alpha_1}{5}}+\gamma_2\,t^{\frac{2}{5}}\right)},\\
\\
\sigma_2^2\,=\,\dfrac{2\,\gamma_1\,(3+\sqrt{19})\,x^{\frac{\alpha_1}{5}}+2\,\gamma_2\,(1+\sqrt{19})\,t^{\frac{2}{5}}}{
30\,t\,\left(\gamma_1\,x^{\frac{\alpha_1}{5}}+\gamma_2\,t^{\frac{2}{5}}\right)},\\
\\
\sigma_3^3\,=\,-\dfrac{\gamma_1\,(18+\sqrt{19})\,x^{\frac{\alpha_1}{5}}+2\,\gamma_2\,(10+\sqrt{19})\,t^{\frac{2}{5}}}{
30\,t\,\left(\gamma_1\,x^{\frac{\alpha_1}{5}}+\gamma_2\,t^{\frac{2}{5}}\right)},
  \end{array}
\right.
\end{equation}

The shear scalar is given by
\begin{equation}\label{uu1-9}
  \begin{array}{ll}
\sigma^2\,=\,\dfrac{(103+6\,\sqrt{19})\,\gamma_1^2\,x^{\frac{2\,\alpha_1}{5}}
+2\,(67+4\,\sqrt{19})\,\gamma_1\,\gamma_2\,x^{\frac{\alpha_1}{5}}\,t^{\frac{2}{5}}+(47+2\,\sqrt{19})\,\gamma_2^2\,t^{\frac{4}{5}}}{
300\,t^2\,\left(\gamma_1\,x^{\frac{\alpha_1}{5}}+\gamma_2\,t^{\frac{2}{5}}\right)^2}.
  \end{array}
\end{equation}

The deceleration parameter is determined as \cite{dec1}
\begin{equation}\label{uu1-11}
  \begin{array}{ll}
\mathbf{q}&=-3\,\Theta^2\,\Big(\Theta_{;i}\,u^{i}+\dfrac{1}{3}\,\Theta^2\Big)\\
&=-3\,\Theta^2\,\Bigg[\dfrac{\Theta^2}{3}-
\dfrac{(18+\sqrt{19})\,x^{\frac{\alpha_1}{2}}}{10\,t^2\,\left(\gamma_1\,x^{\frac{\alpha_1}{5}}+\gamma_2\,t^{\frac{2}{5}}\right)}
+\dfrac{2\,\gamma_2\,x^{\frac{\alpha_1}{2}}\,
\left(3\,\gamma_1\,x^{\frac{\alpha_1}{5}}+5\,\gamma_2\,t^{\frac{2}{5}}\right)}{25\,t^{\frac{8}{5}}\,\left(\gamma_1\,x^{\frac{\alpha_1}{5}}+\gamma_2\,t^{\frac{2}{5}}\right)^3}\Big)\Bigg].
  \end{array}
\end{equation}
\begin{center}
\begin{figure*}[thbp]
\begin{tabular}{rl}
\includegraphics[width=5.5cm]{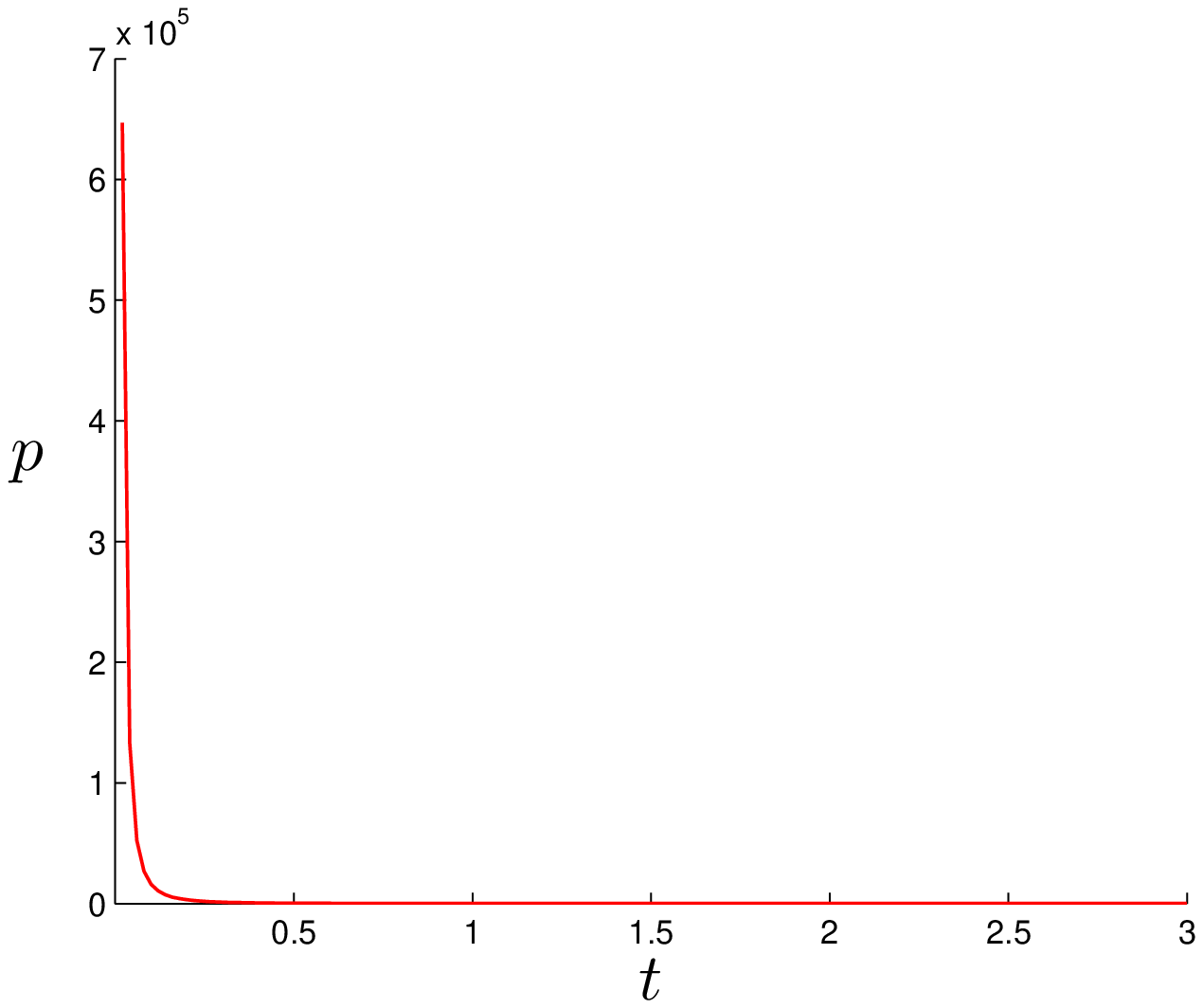}
\includegraphics[width=5.5cm]{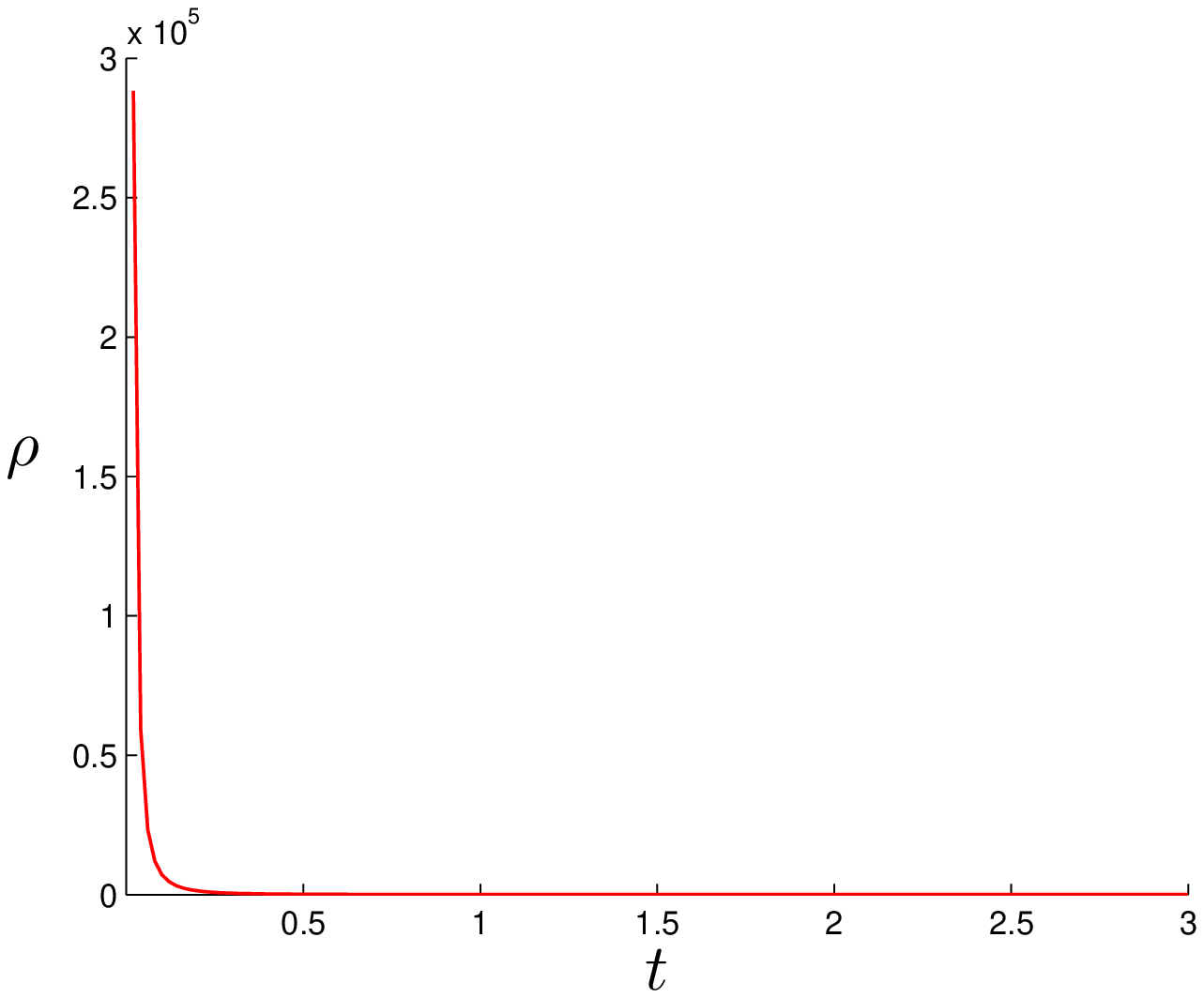}
\includegraphics[width=5.5cm]{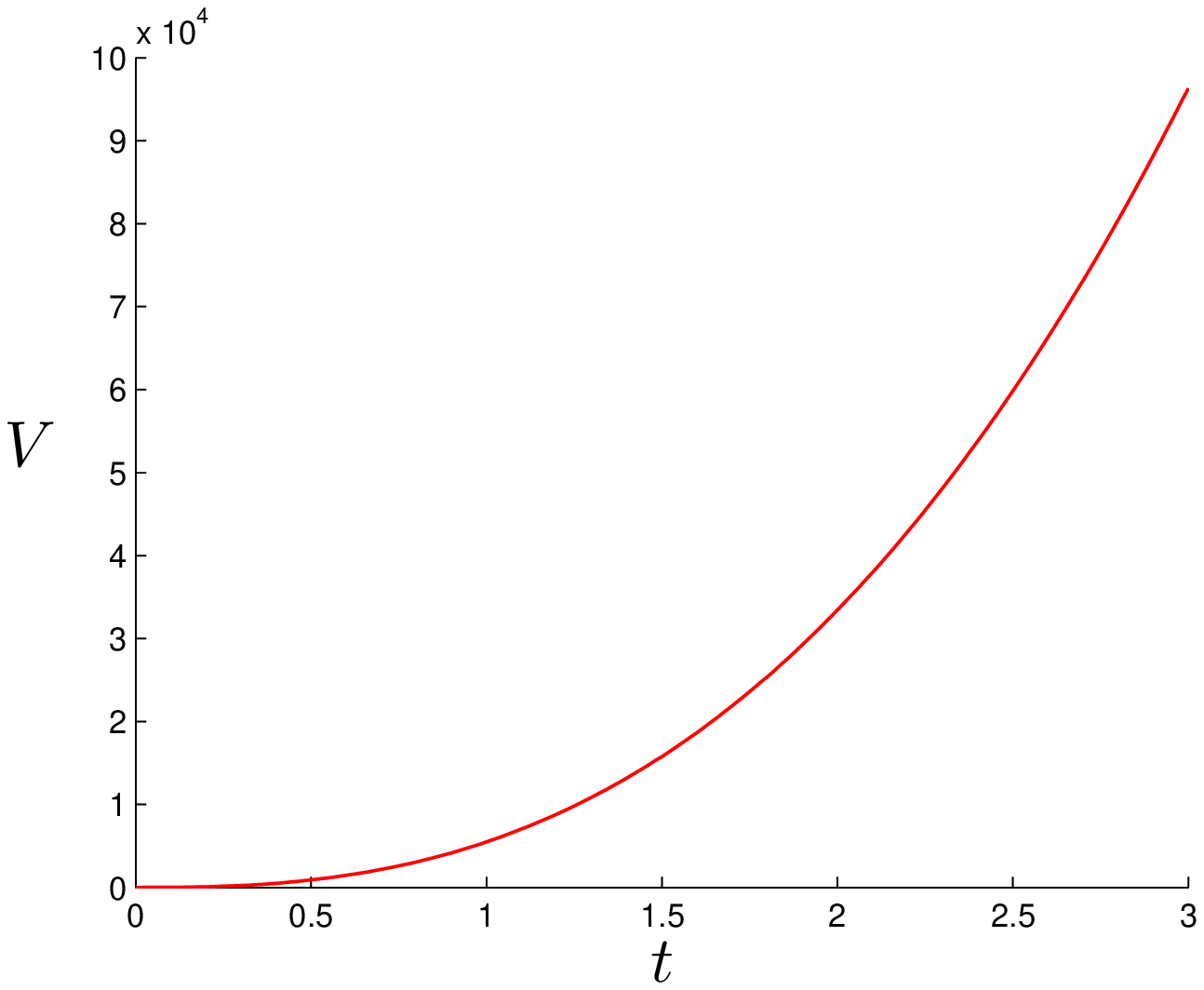}\\
\end{tabular}
\caption{Variation of pressure (left panel), energy density (middle panel) and
volume (right panel) versus time for model (\ref{s1}):}.
\end{figure*}
\end{center}
For model (\ref{s1}), the variation of pressure, energy density and volume are presented in Fig. 1. For 
the large $t$, the pressure and energy density approached towards zero and volume diversed towards infinity i. e. 
$p \rightarrow 0 $, $\rho \rightarrow 0$ and $V \rightarrow \infty$ when $t \rightarrow \infty$. It has singularity 
at $t = 0$. In general the model has point type singularity at $t = 0$ because the scale factor and volume vanish at 
initial epoch. From eq.(\ref{uu1-11}), it is evident that deceleration parameter is negative and 
decreasing function of time and hence the cosmic expansion is driven by big bang impulse. \\ 

\textbf{For the Model (\ref{s3}):}\\

The expressions of $p$ and $\rho$  for the model (\ref{s3}), are
given by:
\begin{equation}\label{uu2-1}
  \begin{array}{ll}
p(x,t)=\dfrac{K_1+
(\kappa+3\,\kappa\,\beta_2+6\,\lambda\,\beta_2)\,(2+3\,\beta_2+\beta_2^2)\,\beta_2\,\alpha_1^2\,t^{2\,\beta_2}\,\exp[2\,\tilde{a}\,x]}{
(\kappa^2+6\,\kappa\,\lambda+8\,\lambda^2)\,\Big[\beta_2\,\alpha_1^2\,(2+\beta_2)\,t^{2\,\beta_2}\,\exp[2\,\tilde{a}\,x]-\tilde{a}^2\Big]},
  \end{array}
\end{equation}

\begin{equation}\label{uu2-2}
  \begin{array}{ll}
\rho(x,t)=\dfrac{\Big[3(1+3\beta_2)\kappa+(8+6\beta_2)\lambda\Big](2+3\beta_2+\beta_2^2)\beta_2\alpha_1^2\,t^{2\,\beta_2}\,\exp[2\,\tilde{a}\,x]
-K_2}{
(\kappa^2+6\,\kappa\,\lambda+8\,\lambda^2)\,\Big[\beta_2\,\alpha_1^2\,(2+\beta_2)\,t^{2\,\beta_2}\,\exp[2\,\tilde{a}\,x]-\tilde{a}^2\Big]},
  \end{array}
\end{equation}
where $\beta_2$, $\tilde{a}$ and $\lambda$ are an arbitrary
constants while $\kappa=8\,\pi$,
$K_1=(\kappa+\kappa\,\beta_2+2\,\lambda\,\beta_2)\,(1+3\,\beta_2)\,\tilde{a}^2$
and
$K_2=\Big[(3+\beta_2)\kappa+(8+2\beta_2)\lambda\Big](1+3\,\beta_2)\tilde{a}^2$.
The volume element is given by
\begin{equation}  \label{uu2-4}
V=\alpha_1\,\beta_1\,\tilde{\gamma}_1^2\,\Big[\tilde{a}^2-\beta_2\,\alpha_1^2\,(2+\beta_2)\,t^{2\,\beta_2}\,\exp[2\,\tilde{a}\,x]\Big]\,
\,t^{3+\beta_2}\,\exp[2\,\tilde{a}\,x],
\end{equation}
where $\beta_1$ and $\tilde{\gamma}_1$ are an arbitrary constants.\\
The expansion scalar, which determines the volume behavior of the
fluid, is read as
\begin{equation}\label{uu2-5}
  \begin{array}{ll}
\Theta=\dfrac{\tilde{a}^2\,(3+\beta_2)-3\,\beta_2\,\alpha_1^2\,(2+3\,\beta_2+\beta_2^2)\,t^{2\,\beta_2}\,\exp[2\,\tilde{a}\,x]}{
t\,\Big[\tilde{a}^2-\beta_2\,\alpha_1^2\,(2+\beta_2)\,t^{2\,\beta_2}\,\exp[2\,\tilde{a}\,x]\Big]},
  \end{array}
\end{equation}
The non-vanishing components of the shear tensor, $\sigma_i^j$, are read as
\begin{equation}\label{uu2-6}
\left\{
  \begin{array}{ll}
\sigma_0^0\,=\,0,\,\,\,\,\,\sigma_1^1\,=\,
\dfrac{2\,\tilde{a}^2\,\beta_2}{3\,t\,\left(\tilde{a}^2-\beta_2\,(\beta_2+2)\,\alpha_1^2\,t^{2\,\beta_2}\,\exp[2\,\tilde{a}\,x]\right)},\,\,\,\,\,
\sigma_2^2\,=\,\sigma_1^1,\,\,\,\,\,\sigma_3^3\,=\,-2\,\sigma_1^1.
  \end{array}
\right.
\end{equation}
The shear scalar is given by
\begin{equation}\label{uu2-9}
  \begin{array}{ll}
\sigma^2\,=\,\dfrac{4\,\tilde{a}^4\,\beta_2^2}{3\,t^2\,\left(\tilde{a}^2-\beta_2\,(\beta_2+2)\,\alpha_1^2\,t^{2\,\beta_2}\,\exp[2\,\tilde{a}\,x]\right)^2}.
  \end{array}
\end{equation}
The deceleration parameter is obtained as
\begin{equation}\label{uu2-11}
  \begin{array}{ll}
\mathbf{q}=-3\,\Theta^2\,\Big(\Theta_{;i}\,u^{i}+\dfrac{1}{3}\,\Theta^2\Big)\\
\,\,\,\,\,=-3\,\Theta^2\,\Bigg[\dfrac{\Theta^2}{3}+\dfrac{t^{\beta_2-3}}{\tilde{\gamma}_1\,
\Big[\tilde{a}^2-\beta_2\,\alpha_1^2\,(2+\beta_2)\,t^{2\,\beta_2}\,\exp[2\,\tilde{a}\,x]\Big]^2}\,
\Big(\tilde{a}^4\,(\beta_2-3)\\
\\
\,\,\,\,\,\,\,\,\,\,\,\,\,\,\,\,\,\,\,\,
+2\,\tilde{a}^2\,\beta_2\,\alpha_1^2\,(2+\beta_2)\,(3+2\,\beta_2-\beta_2^2)\,t^{2\,\beta_2}
\,\exp[2\,\tilde{a}\,x]+3\,\beta_2^2\,\alpha_1^4\,(2+3\,\beta_2+\beta_2^2)^2\,t^{4\,\beta_2}\,\exp[4\,\tilde{a}\,x]\Bigg)\Bigg].
  \end{array}
\end{equation}

\begin{figure*}[thbp]
\begin{center}
\begin{tabular}{rl}
\includegraphics[width=6.5cm]{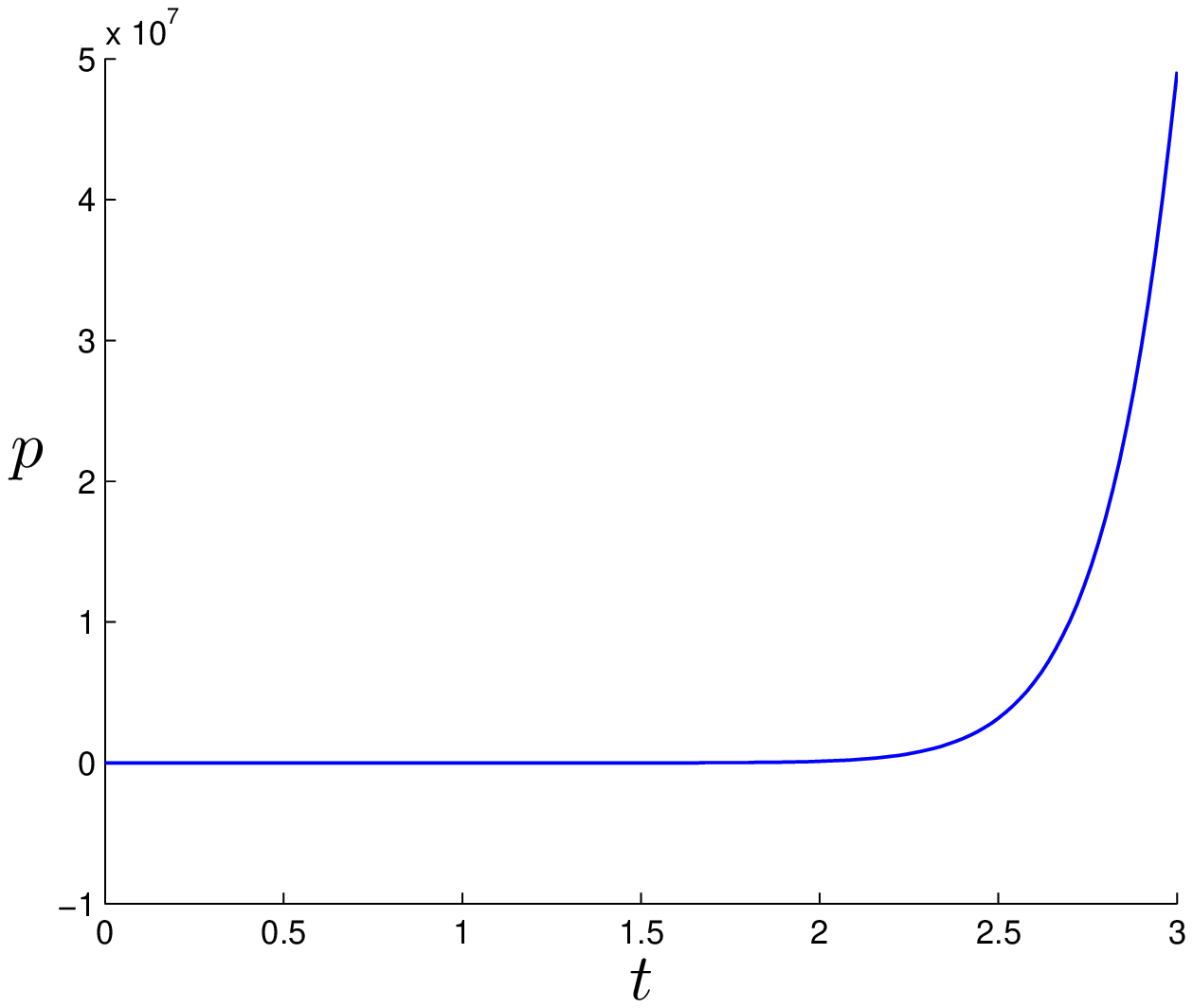}
\includegraphics[width=6.5cm]{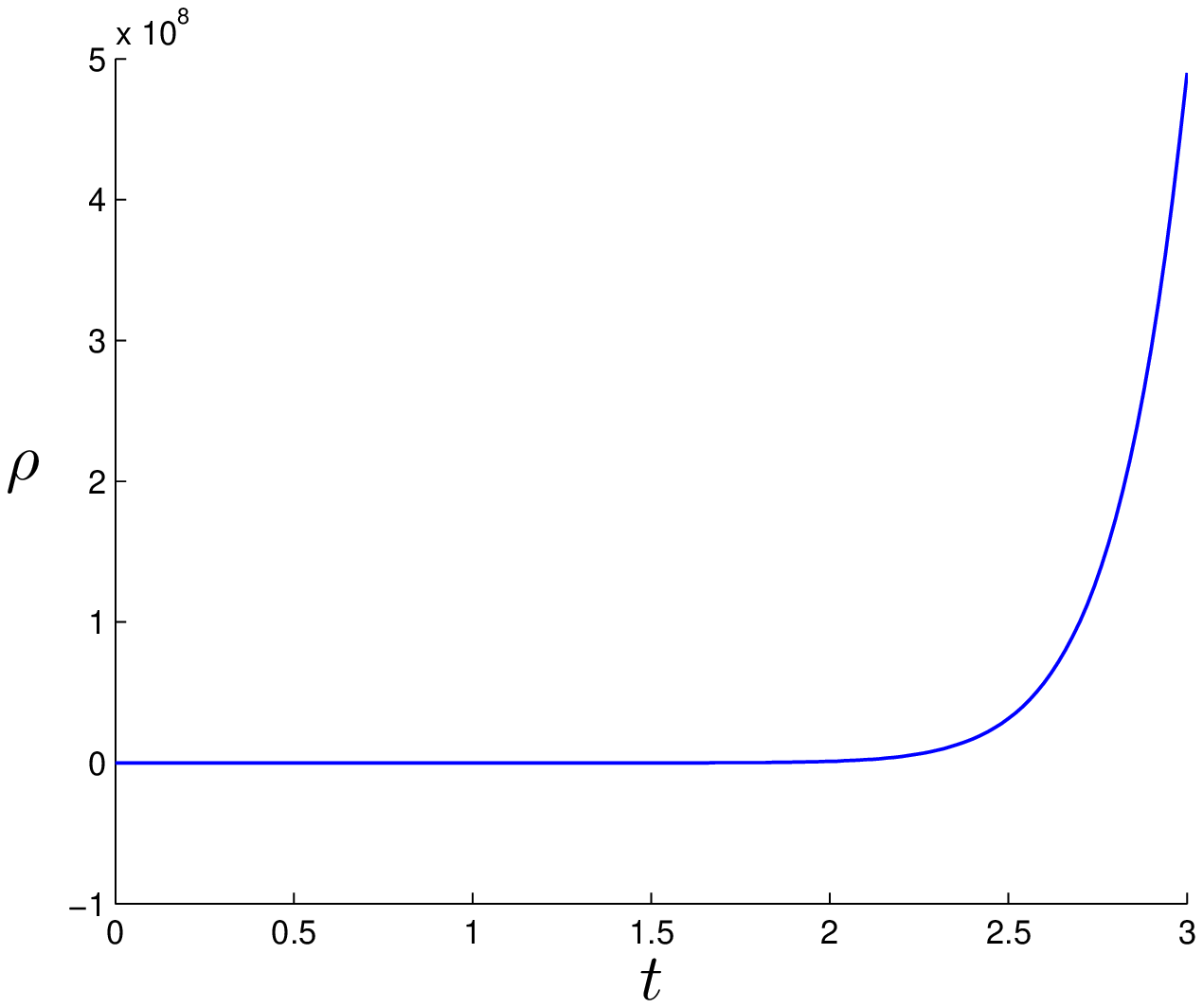}
\end{tabular}
\end{center}
\caption{Variation of pressure (left panel) and energy density (middle panel) versus time for model (\ref{s3}):}.
\end{figure*}
For Physically viable model, one have some certain criterion such as: (i) The energy density should be 
positive and decreases with time. (ii) The Volume of universe increases with increase in time. 
(iii) $\frac{\sigma}{\theta}$ should be vanish at larger time.\\

The Model (\ref{s3}) does not meet criterion (i) as stated above (Fig. 2). Here, in the model (\ref{s3}), 
the energy density increases with time. Therefore this model are not physically interesting and we omit the further 
physical and geometrical analysis of model (\ref{s3}).  

\section {Conclusion}

In this paper, we have investigated an optimal system and invariant solutions of Bianchi
type I space-time in the context of $f\left(R,T\right)$ Gravity. 
In $f(R,T)$ gravity, the cosmic acceleration may arises not only due to geometrical contribution of 
the matter but also depends on matter contents of the universe. In this gravity an extra acceleration is always 
present due to coupling between matter and geometry. We have derived the gravitational field 
equations for the fluid under consideration, corresponding to the $f(R,T)$ gravity models. 
On the basis of optimal systems of symmetries $X^{(1)}$ and $X^{(3)}$, we obtained
two models (\ref{s1}) and (\ref{s3}) respectively. For model (\ref{s1}), we note that at $t = 0$, the spatial volume 
vanishes while other parameter diverge. Thus the model (\ref{s1}) starts expanding with 
big bang singularity at $t = 0$. This singularity is point type because the directional scale factors $A$, $B$ and 
$C$ vanish at $t = 0$. For model (\ref{s3}), we observe that pressure and energy density are increasing function of 
time which is not in favour of standard cosmological model governed by big bang cosmology. Therefore, 
it is interesting to note that the present work deals singular as well as non singular model of universe under the 
specification of symmetries $X^{(1)}$ and $X^{(3)}$ respectively. However, to have consistency with observational data 
(from SN Ia, BAO and CMB) of standard cosmology, the model (\ref{s1}) is suitable model of present universe. In other 
words, the solution presented here has potential to understand the features of observed universe. We also note that 
$\frac{\sigma}{\theta} \rightarrow 0$ at $t\rightarrow \infty$. This means that anisotropy will be died out at larger 
time.\\   


\end{document}